# Using customized GPT to develop prompting proficiency in architectural AI-generated images

Juan David Salazar Rodriguez[1], Sam Conrad Joyce[1], Julfendi[2]

[1] META Design lab, Architecture and Sustainable Design Pillar, Singapore University of Technology and Design, Singapore, Singapore
[2] Engineering Product Development Pillar, Singapore University of Technology and Design, Singapore, Singapore
`1008372@mymail.sutd.edu.sg`, `sam_joyce@sutd.edu.sg`,
`julfendi_julfendi@mymail.sutd.edu.sg`

**Abstract.** This research investigates the use of customized GPT models to enhance prompting proficiency among architecture students when generating AI-driven images. Prompt engineering is increasingly essential in architectural education due to the widespread adoption of generative AI tools. This study utilized a mixed-methods experimental design involving architecture students divided into three distinct groups: a control group receiving no structured support, a second group provided with structured prompting guides, and a third group supported by both structured guides and interactive AI personas. Students engaged in reverse engineering tasks, first guessing provided image prompts and then generating their own prompts, aiming to boost critical thinking and prompting skills. Variables examined included time spent prompting, word count, prompt similarity, and concreteness. Quantitative analysis involved correlation assessments between these variables and a one-way ANOVA to evaluate differences across groups. While several correlations showed meaningful relationships, not all were statistically significant. ANOVA results indicated statistically significant improvements in word count, similarity, and concreteness, especially in the group supported by AI personas and structured prompting guides. Qualitative feedback complemented these findings, revealing enhanced confidence and critical thinking skills in students. These results suggest tailored GPT interactions substantially improve students' ability to communicate architectural concepts clearly and effectively.

## 1 Introduction

Artificial intelligence (AI) has emerged as a transformative force in the field of architecture, fundamentally altering how designers approach creativity, sustainability, and innovation. AI tools, including generative design algorithms, deep learning models, and co-creative systems, are now integral to architectural practice, empowering designers to explore innovative forms, optimize structural efficiency, and enhance environmental sustainability (Wan, 2023; Oxman, 2017; Debrah, Chan, & Darko, 2022). While these advancements offer tremendous potential, they also introduce new challenges, particularly in the context of architectural education. The effective utilization of AI in design demands a sophisticated understanding of how to communicate design ideas through precise and contextually appropriate prompts (Kim, Maher, & Siddiqui, 2021; Liu & Chilton, 2022).

As AI becomes more deeply embedded in architectural education, it is essential for students to develop skills in prompt engineering and critical thinking to fully harness AI's capabilities (Walter, 2024). However, the steep learning curve associated with mastering these tools often results in suboptimal design outputs, particularly when students lack sufficient guidance on how to articulate their design intentions effectively (Ceylan et al., 2024). This challenge is exacerbated in online and hybrid learning environments, where the shift to digital tools has highlighted the need for a more structured approach to integrating AI into design education (Życzkowska & Urbanowicz, 2019).

To address this gap in architectural education, this research introduces a structured approach to AI prompting. The study investigates how architectural students with the help of a prompting guide using a customized GPT that includes prompting tips, tasks, feedback and AI personas, interact during the image AI-generation design process and examines how these interactions influence the quality of the prompts generated. The primary research questions guiding this study are:

1. How a crafted guide that includes parameters in the architectural field (subject, medium, environment, materials, architectural style, lighting, colors, environmental and social aspects) improves architects abilities for prompting?
2. How do AI personas representing different construction industry roles influence the vocabulary of students' prompts?
3. What metrics (time spent, word count and concreteness) most accurately correlate to successful prompting in architectural design?

The findings from this study are expected to significantly contribute to the field of architectural education by showing tools that educators can use to develop more targeted teaching strategies such as using a customized GPT for learning prompting skills. This research aims to bridge the gap between technology and pedagogy in architecture, providing valuable insights that will improve the integration of AI in architectural curricula (Figoli, Rampino, & Mattioli, 2022; Rezwana & Maher, 2023; As, Pal, & Basu, 2018).

## 2 Literature review

### 2.1 AI in architecture

Artificial Intelligence (AI) is reshaping architecture by augmenting creativity, optimizing design processes, and addressing sustainability challenges. AI tools such as generative design algorithms, deep learning models, and co-creative systems enable architects to explore innovative forms and optimize the functionality and environmental performance of designs (Wan, 2023; Oxman, 2017; Rafsanjani & Nabizadeh, 2023). These advancements have revolutionized how architects approach conceptual design, integrating computational techniques like evolutionary algorithms to explore solutions during early-stage design when parameters remain undefined (Chew et al., 2024; Saadi & Yang, 2023). This flexibility allows the generation of a wide range of design alternatives, empowering architects to break traditional creative boundaries.

Generative design exemplifies this transformation, enabling the automated exploration of diverse design solutions through computational algorithms. Platforms such as Autodesk's generative design software integrate with Building Information Modeling (BIM), offering real-time optimization capabilities that enhance efficiency across industries, including architecture and construction (Cudzik & Radziszewski, 2018; Patel et al., 2024). These tools contribute to sustainability by reducing material waste, optimizing energy performance, and integrating environmental factors into architectural solutions, such as facade design and energy-efficient skyscrapers (Bölek et al., 2023; Castro Pena et al., 2021).

Beyond design generation, AI supports architectural preservation by aiding in the reconstruction of historical sites and digitalizing cultural heritage. This dual focus on innovation and tradition illustrates the breadth of AI's impact, blending creative exploration with practical applications (Bölek et al., 2023). However, as AI increasingly permeates architectural practice, challenges arise in balancing automation with human creativity and expertise. Rafsanjani and Nabizadeh (2023) advocate for human-centered AI systems that align with architectural decision-making processes, emphasizing the collaboration between human intuition and machine precision (Davis et al., 2021).

AI's transformative potential extends into architectural education, where it introduces new complexities and opportunities. Developing skills in AI tools, particularly in prompt engineering and critical thinking, is crucial for students to effectively communicate design intentions and achieve desired outcomes (Walter, 2024; Życzkowska & Urbanowicz, 2019). The steep learning curve associated with AI tools often leads to challenges in articulating design concepts, highlighting the need for structured teaching strategies. This issue is amplified in digital and hybrid learning environments, underscoring the necessity of aligning AI capabilities with pedagogical practices (Ceylan et al., 2024).

Emerging AI applications, such as text-to-image generation platforms (e.g., DALL-E 2, MidJourney), offer architects tools to visualize complex ideas and facilitate client communication. These tools exemplify AI's creative potential, yet they demand careful refinement to align outputs with professional standards and workflow requirements (Hanafy, 2023; Zhang et al., 2023). The role of AI as both a design assistant and a co-creator introduces a dynamic interplay between technology and human ingenuity, necessitating continued research and innovation to optimize its integration into architectural practice (Rane, 2024). As this study explores, integrating structured prompting guides and AI personas tailored to architectural practice could significantly enhance the creative process. These approaches aim to bridge gaps in current AI usage, fostering improved communication, efficiency, and design quality within architectural education and practice. This alignment of technology with human creativity reinforces AI's potential to redefine the future of architecture.

**2.2 Architectural education and digital pedagogy**

The integration of advanced technologies such as Artificial Intelligence (AI), digital tools, and immersive platforms is redefining architectural education, compelling educators to rethink traditional pedagogical approaches. These innovations offer unprecedented opportunities for enhancing creativity and problem-solving but also necessitate a balance between traditional skills and emerging technological competencies. As noted by Kara (2015), foundational manual skills like hand drawing and physical modeling remain crucial for understanding spatial and tectonic principles. These skills serve as a base for integrating advanced digital tools, ensuring that technology complements rather than overshadows core architectural principles (Xiang et al., 2020; Vrontissi, 2015).

AI presents unique potential in architectural education, offering tools like generative design and text-to-image platforms that enable students to explore complex design problems and innovative solutions. However, utilizing these tools effectively requires a sophisticated understanding of prompt engineering, a challenge addressed in this study through a crafted prompting guide and AI personas. Walter (2024) emphasizes the importance of structured strategies in teaching students to articulate design intentions, particularly in environments where digital tools dominate. The shift toward digital education has also transformed traditional critique methods in architectural studios. Historically central to design education, critique sessions often fail to adapt to modern pedagogical needs, leading to student dissatisfaction (Utaberta et al., 2013). This study addresses this gap by introducing iterative feedback mechanisms through AI personas, aligning critique processes with collaborative and student-centered digital environments. This approach aims to improve the pedagogical alignment of critiques while enhancing the quality of design outcomes.

The role of physical studio spaces in architectural education remains significant despite the increasing dominance of digital tools. Corazzo (2019) and McClean (2009) argue that studio environments shape students' professional identities and critical thinking skills. However, in hybrid and online settings, maintaining this dynamic requires innovative approaches, such as those employed in this study, where participants engage with AI-guided prompts in collaborative settings to simulate studio interactions. Sustainability education also plays a critical role in modern architectural curricula, reflecting broader environmental and social imperatives. Models integrating parametric design and sustainability, as proposed by Hsu and Ou (2022), align with this study's emphasis on embedding contextual considerations like materials, environmental factors, and cultural narratives into AI-generated design processes. Workshops and hands-on experiences have proven effective in fostering creativity and interdisciplinary collaboration in architecture education. By structuring tasks that involve crafting and refining AI-generated prompts, this study mirrors the benefits observed in prior research, where interactive activities encouraged practical skill development and critical engagement (Karslı & Özker, 2014; Arslan & Dazkir, 2017).

## 2.3 Prompting engineering

Prompting engineering, the art and science of crafting effective instructions for AI systems, has emerged as a critical skill in maximizing the potential of generative AI tools. This study situates prompting engineering within the architectural domain, where precision and creativity are paramount. As the effective use of text-to-image AI tools hinges on the quality of prompts, the research emphasizes the development of structured prompting techniques and the integration of AI personas to guide architectural students. This approach aligns with Cain's (2024) assertion that prompt literacy is essential for accessing the full capabilities of large language models (LLMs) and fostering innovative problem-solving. In the architectural context, prompt engineering is more than a technical skill; it is a creative and collaborative process that influences the quality and relevance of AI-generated outputs. Hutson and Robertson (2023) found that successful integration of AI tools into a 3D design course significantly enhanced student creativity, but outcomes depended heavily on the clarity and specificity of the prompts. This underscores the importance of structured teaching methodologies, such as those employed in this study, which uses a crafted prompting guide to introduce students to key components of effective prompts, including subject, medium, environment, materials, and style.

The collaborative potential of prompting is explored through the integration of AI personas—simulated experts representing roles such as sustainability consultants, landscape architects, and interior designers. Han et al. (2024) highlight the benefits of human-AI collaboration during creative tasks, showing that co-prompting strategies can foster innovation and refine design outcomes. In this study, AI personas not only provide tailored vocabulary suggestions but also simulate interdisciplinary collaboration, offering students a practical understanding of the diverse considerations involved in architectural design. Prompting engineering also serves as a catalyst for self-directed learning, empowering students to iterate and refine their prompts independently. Garg and Rajendran (2024) demonstrate that structured prompting frameworks enhance learners' ability to tackle complex tasks, fostering both confidence and competence. This study builds on these findings by incorporating iterative feedback loops, where students receive metrics-based evaluations of their prompts—such as word count, time spent, and abstraction level—alongside constructive guidance for improvement.

The creative dimensions of prompting are exemplified in artistic applications, where prompts are not just instructions but integral elements of the creative process. Chang et al. (2023) explore how prompt engineering enables artists to navigate text-to-image platforms as a new medium for expression. Similarly, in architectural education, the iterative refinement of prompts is positioned as both a learning exercise and a creative exploration, bridging the gap between abstract ideas and concrete design representations (Almeda et al., 2024). The social and rhetorical aspects of prompting further highlight its multidimensional role. Goloujeh, Sullivan, and Magerko (2024) emphasize the communal nature of prompt development within AI user communities, where shared knowledge and collaborative practices enhance the quality of interactions with generative models. This aligns with the study's methodology, which introduces structured group tasks and AI personas to simulate professional collaboration, encouraging students to adopt diverse perspectives in their design processes.

Ethical considerations are central to prompting engineering, particularly in educational settings. Hutson and Cotroneo (2023) discuss the challenges of ensuring responsible use of AI tools, advocating for pedagogical strategies that balance innovation with ethical awareness. This study incorporates these principles by emphasizing the contextual and environmental dimensions of architectural prompts, teaching students to integrate sustainability and cultural sensitivity into their AI-mediated designs. Finally, as AI technologies continue to evolve, the adaptability of prompting techniques becomes increasingly important. Lo (2023) and Schleith et al. (2022) advocate for continuous refinement of prompting strategies to align with advances in AI capabilities. This study reflects this need for adaptability by progressively increasing the complexity of prompting tasks, introducing additional parameters such as lighting, resolution, and architectural styles as students advance through the modules.

## 3 Research objectives

This study explores the integration of text-to-image AI tools within the architectural design process, particularly analyzing how architectural students interact with these tools to produce visual representations. This research delves into the effectiveness of a crafted prompting guide for image AI generated content in the architectural field and AI-generated personas, which simulate the expertise of professionals such as interior designers, landscape architects, and sustainability consultants. These personas are used to guide students in crafting more precise and effective prompts. The study's methodology systematically evaluates the impact of different prompt parameters, comparing quantitatively the design outputs between different groups, a control group without any explanation for prompting, another group that has explanations of the prompting and an experimental group that receives assistance from these AI personas. The aim is to assess how these tools and guided interactions can enhance or alter the creative process in architectural education.

### 3.1 Primary objectives

The primary objective of this research is to assess the effectiveness of text-to-image AI tools in the architectural design process and to determine how the introduction of a crafted prompting guide and the use of AI-generated personas influences the prompting skills and design outcomes of architectural students.

### 3.2 Secondary objectives

- To analyze the relationship between different prompt parameters (word count, time spent on prompting, accuracy, and abstractness) of the generated images.

- To compare the performance and learning outcomes between students who work independently without any guide and those who are assisted by AI personas and a prompting guide.

- To evaluate the pedagogical potential of integrating AI tools in architectural education, particularly in teaching effective prompting techniques.

## 4 Methods

### 4.1 Experimental Design

#### 4.1.1 Participants

The study involved 36 architectural students of differential educational levels (bachelor, masters, doctoral), selected through a stratified random sampling method to ensure a diverse representation of skill levels and backgrounds. Participants were randomly assigned to one of three groups:

- Experimental Group 1 (Group 1): 12 participants crafted AI generated images with a tailored prompting guide and assistance from AI-generated personas designed to emulate the expertise of various construction industry professionals.

- Experimental Group 2 (Group 2): 12 participants created AI generated images with a tailored guide to teach them prompt skills but no AI personas assistance.

- Control Group (Group 3): 12 participants crafted AI-generated images without a guide or external assistance.

#### 4.1.2 Tools

The study employed a variety of tools to support its methodology. A customized GPT model played an important role in simulating AI personas that provide domain-specific guidance, enriching participants' understanding of architectural concepts and prompting strategies. DALL·E was utilized to generate text-to-image outputs, allowing participants to visualize their prompts and refine them iteratively. Pre- and post-experiment surveys were conducted through Microsoft Forms, incorporating both Likert-scale and open-ended questions to evaluate participants' experiences and learning outcomes.

### 4.1.3 Research Procedure

The procedure for this research included preliminary assessments, the main experimental tasks, data collection, and post-experiment evaluations.

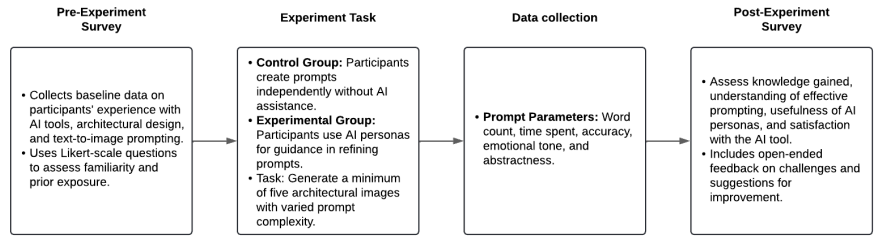

Figure 1. Research procedure.

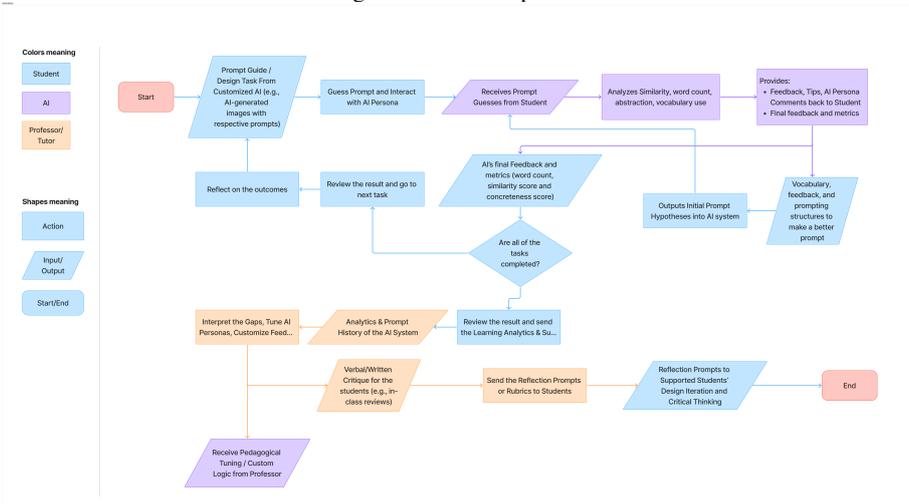

Figure 2: Workflow of Student-AI-Professor Interaction in Prompt-Based Architectural Learning

Figure 2 represents a comprehensive workflow depicting interactions between students, the customized AI system, and the professor/tutor. Initially, students receive a prompt guide or design task generated by a customized AI system, including AI-generated images with respective prompts. Students then engage in the learning task by guessing prompts and interacting with AI personas. The AI system receives these guesses and analyzes them based on various criteria, such as similarity to model prompts, word count, abstraction, and vocabulary use. Subsequently, the AI system provides students with detailed feedback, including similarity scores, concreteness metrics, vocabulary suggestions, and structural tips to enhance prompting skills. Students reflect on outcomes, revising their prompts based on this feedback, and progress iteratively through a sequence of tasks until all designated activities are completed. Once all tasks are finished, the system outputs initial prompt hypotheses, incorporating all student inputs. The professor/tutor then receives an analytics summary and prompt history from the AI system, interpreting gaps in student learning, refining AI personas, and customizing feedback mechanisms. This iterative pedagogical tuning ensures alignment with learning objectives. Finally, the professor sends reflection prompts or rubrics back to the students to foster design iteration and critical thinking, effectively closing the loop and enriching the overall learning experience.

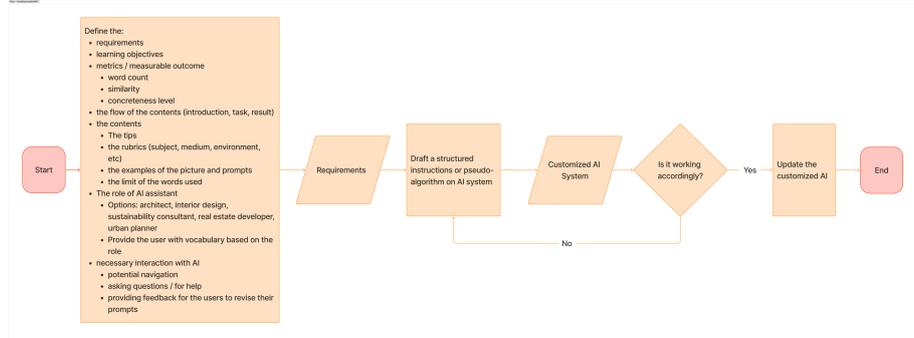

Figure 3: Process for Customizing and Updating the AI System

Figure 3 illustrates the systematic process for customizing and refining the AI system used in this educational research. The customization begins by clearly defining requirements, including explicit learning objectives, measurable outcomes such as word count, similarity, concreteness, and structured content elements like instructional tips, rubrics (subject, medium, environment), example prompts and images, and constraints such as word limits. It also defines the roles for AI assistants, which represent various architectural and urban planning professionals, including architects, interior designers, sustainability consultants, real estate developers, and urban planners. These roles are intended to enrich the vocabulary and contextual relevance provided to students. Following requirement definition, structured instructions or a pseudo-algorithm are drafted for integration into the AI system. Once integrated, the customized AI system undergoes testing to evaluate its alignment with intended outcomes. If the AI operates as expected, updates are formally implemented into the AI system. If discrepancies or suboptimal results are observed, the drafting process is revisited, and adjustments are made iteratively. This ensures continuous improvement and alignment of the AI system with pedagogical goals, resulting in a tailored educational tool optimized for architectural prompting tasks.

### 4.1.3.1 Pre experiment survey

The pre-experiment survey was designed to establish a baseline understanding of each participant's background and prior experience with AI tools in architectural design. The survey collected demographic information, including age, gender, level of study (bachelor, master, doctoral) and professional experience in architecture or related fields. This demographic data ensured a diverse representation of participants and facilitate stratified random sampling.

In addition to demographics, the survey assessed participants' general familiarity and experience with AI tools through a series of Likert-scale questions. These questions evaluated the frequency of AI tool usage, confidence in learning new AI technologies, and the extent to which participants seek out AI tools to enhance their work processes. A specific focus was placed on prior exposure to text-to-image AI tools, which are directly relevant to the experiment's tasks. Participants rated their familiarity with these tools, their ability to create effective prompts, satisfaction with past AI-generated outcomes, and confidence in improving images through prompt refinement.

**Section A: Demographic Information**

| Question | Answer Options |
|---|---|
| Age | [Open Response] |
| Gender | - Male<br>- Female<br>- Non-binary<br>- Prefer not to say |
| Level of Study | - Bachelor<br>- Master<br>- Doctoral |
| Professional Experience in Architecture or Related Fields | [Open Response] |

**Section B: Experience with AI Tools**

*Please rate your agreement with the following statements on a scale from 1 to 5, where 1 means "Not at All" and 5 means "Extremely."*

| Question Number | Statement | Rating Scale (1-5) |
|---|---|---|
| 1 | I have overall experience with AI tools. | 1 - 2 - 3 - 4 - 5 |
| 2 | I frequently use AI tools in my academic or professional work. | 1 - 2 - 3 - 4 - 5 |
| 3 | I am confident in my ability to learn and adapt to new AI tools. | 1 - 2 - 3 - 4 - 5 |
| 4 | I actively seek out AI tools to enhance my work processes. | 1 - 2 - 3 - 4 - 5 |

**Section C: Prior Exposure to Text-to-Image AI Tools**

| Question Number | Statement | Rating Scale (1-5) |
|---|---|---|
| 5 | I am familiar with text-to-image AI tools. | 1 - 2 - 3 - 4 - 5 |
| 6 | I have used text-to-image AI tools for generating visual content. | 1 - 2 - 3 - 4 - 5 |
| 7 | I can create effective prompts for text-to-image AI tools. | 1 - 2 - 3 - 4 - 5 |
| 8 | I have been satisfied with the outcomes generated by text-to-image AI tools in the past. | 1 - 2 - 3 - 4 - 5 |
| 9 | I am confident in my ability to improve AI-generated images through prompt refinement. | 1 - 2 - 3 - 4 - 5 |

Table 1. Pre survey questions

### 4.1.3.2 Experiment task

In the initial phase, users were introduced to the concept of prompting within the architectural context. Prompting was framed as a critical skill that allowed users to describe their design visions clearly and effectively, guiding AI tools in generating accurate visual outputs. The introduction explained that a strong prompt typically includes a subject, medium, environment, and style, providing a foundation for students to build upon. A sample prompt was presented to illustrate this point: "A contemporary home, built primarily from timber and glass in a watercolor painting, situated on a sloping hill overlooking a forested lake." An image generated from this prompt was displayed to reinforce the connection between the textual description and its visual outcome. Users were then invited to either proceed to the next session or explore additional examples to deepen their understanding.

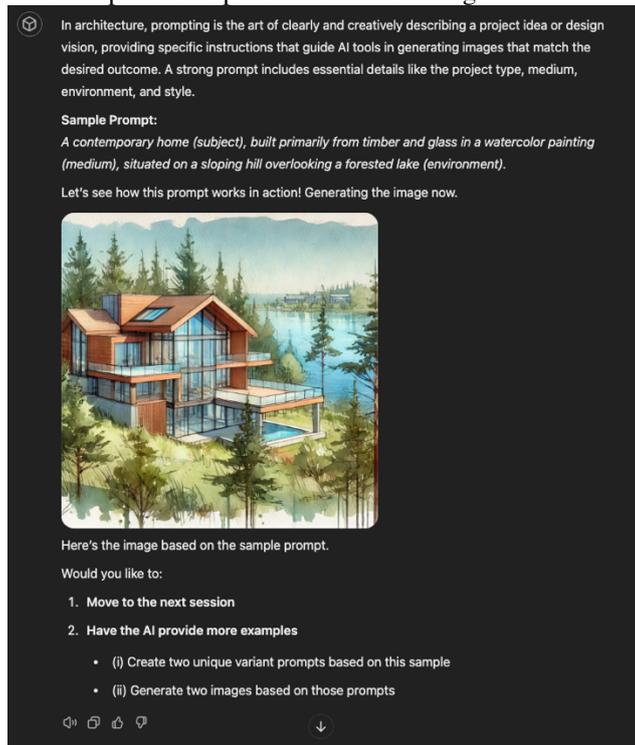

Figure 4. Archiprompt introduction to prompting for image gen-AI in architecture.

The subsequent phase introduced core prompting techniques through structured tips. Each tip focused on a key element of a well-crafted prompt, starting with the fundamental components: subject, medium, and environment. For instance, users were presented with the example: "A modern library rendered in pencil sketches, located in a bustling urban plaza." Alongside this prompt, the corresponding image was displayed to demonstrate how these components influence the visual output. Users were offered the choice to either begin practicing or request further explanation of the example, ensuring that the learning process is both flexible and thorough.

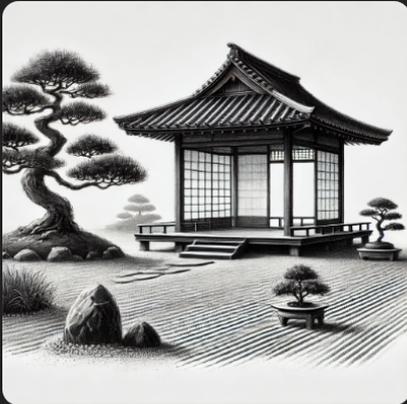

Figure 5. Archiprompt tips section.

The task section builds on these tips by offering hands-on exercises where users crafted prompts themselves. For each task, users were presented with an architectural image and asked to compose a prompt that could describe it. These prompts were structured to follow a specific pattern: "A [subject] rendered in [medium] set in [environment]." For example, an image of a tropical villa might correspond to the hidden prompt: "A tropical villa rendered in watercolor located on a coastal cliff." Users were encouraged to create prompts that adhere to word count constraints (10–20 words initially, with increasing complexity in later tasks) while maintaining clarity and relevance. The system evaluated the user's submission and provides iterative feedback, guiding them to refine their prompt until it aligns with the intended structure and visual context.

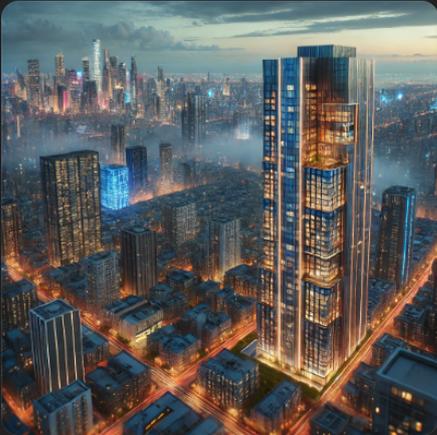

Figure 6. Archiprompt task section.

Once the task was completed, the system transitioned to the result section, where the user's performance was evaluated. The evaluation included an analysis of word count, a comparison of the user's prompt with the original, and a similarity percentage. Additionally, the concreteness of the prompt was assessed using Brysbaert's concreteness ratings, with scores categorized as high abstractness, moderate abstractness/concreteness, or high concreteness. This comprehensive feedback was presented to the user, along with a report summarizing their progress and areas for improvement. Users were then invited to either ask questions or advance to the next tip, ensuring a continuous and supportive learning process.

Figure 7. Archiprompt prompt feedback.

Figure 8. Archiprompt results section.

As users progressed through the modules, they encountered increasingly complex prompting tasks. Beyond the initial focus on subject, medium, and environment, later modules introduced additional dimensions such as materials, architectural style, lighting, colors, and resolution. For example, the second module involved prompts like: "A modern house from bamboo, recycled steel, and glass in a suburban area rendered in oil painting, featuring a contemporary style with clean lines, open spaces, and green roofs." The third module further challenged users by incorporating elements such as lighting and resolution: "An Indonesian boutique shop in suburban glass and concrete rendered in colored pencil sketches. The render is high resolution at 150 DPI." Each module builds on the previous one, gradually expanding the user's understanding and capability to craft nuanced prompts.

A distinctive feature of this study was the integration of AI personas, which simulate professional expertise in various fields related to architecture. Users could select a persona, such as an architect, interior designer, or sustainability consultant, who provided vocabulary suggestions tailored to the task at hand. For example, if the user chose a sustainability consultant persona, the GPT suggested terms like "green roofs," "recycled materials," or "passive solar design" to enrich the prompt. This feature not only diversified the user's learning experience but also demonstrated the interdisciplinary nature of architectural design.

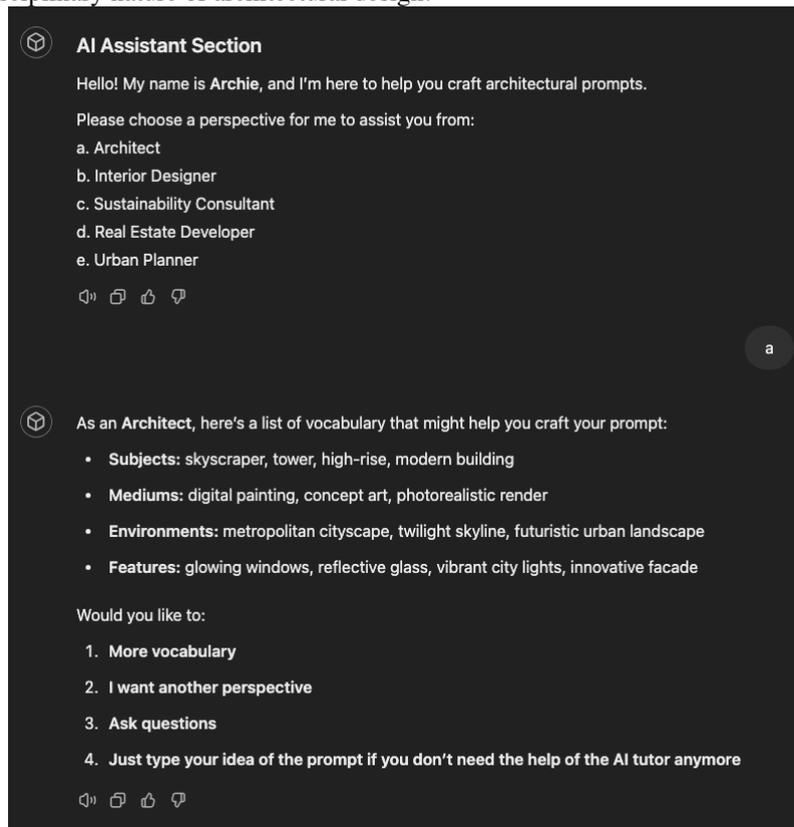

Figure 9. Archiprompt AI personas section.

#### 4.1.3.3 Data collection

Once participants completed their prompts, they submitted them to the AI tool, which generated architectural images based on the provided text and could see the contrast with the original prompt. Each participant was required to generate a minimum of four images during the whole experiment, with prompts that varied in complexity and scope. The generated images were automatically saved and categorized for further analysis. Throughout the task, several key metrics were recorded for each participant:

- **Word Count:** measured the length of the prompts generated by participants. It was an indicator of the detail and complexity of the instructions provided to the AI tool. A higher word count suggested more detailed and potentially more precise prompts. In the experimental group, higher word counts may indicate that AI personas encouraged participants to be more thorough in their descriptions.
- **Time Spent:** amount of time participants took to create each prompt. It reflects the cognitive effort and the thoroughness with which participants approached the task. A shorter time spent in the experimental group might suggest that AI guidance made the process more efficient, helping participants generate prompts more quickly without sacrificing quality.
- **Similarity:** measured how closely the AI-generated images match the participants' intended designs. This is typically rated on a scale, with higher scores indicating better alignment with the intended outcome. Higher accuracy scores in the experimental group would indicate that AI personas helped participants produce prompts that led to more accurate or satisfactory image outputs.
- **Concreteness:** measured the level of conceptual or non-literal thinking in the prompts, typically on a scale from abstract (low concreteness) to highly concreteness. Higher abstractness levels in the experimental group might indicate that AI personas encouraged participants to think more creatively or conceptually, possibly leading to more innovative designs.

These variables were crucial for understanding how different aspects of the prompt affect the quality and relevance of the generated images. They also provide a basis for comparing the performance of the control and experimental groups. After submitting their prompts and generating images, participants in the experimental group received immediate feedback showing them the metrics recorded for the task, see Figure 6. Participants were encouraged to reflect on this feedback and consider how they might improved their prompting techniques in future tasks.

#### 4.1.3.4 Post-experiment survey

Following the completion of the experimental tasks, the post-experiment survey was administered to evaluate the impact of the interventions on participants' prompting skills, confidence with AI tools, and overall experience during the study. The survey aimed to measure any changes in self-reported abilities and perceptions compared to the baseline data collected in the pre-experiment survey. The post-experiment survey included Likert-scale questions that assessed participants' confidence in creating effective prompts, understanding of prompt structuring, and ability to improve AI-generated images through refinement. Additional questions evaluated perceived improvements in prompting skills, enhanced understanding of AI tools in architectural design, and the likelihood of future usage of text-to-image AI tools in academic or professional work.

For participants in the experimental groups, the survey specifically examined the effectiveness of the tailored prompting guide and AI personas. Questions focused on how helpful these resources were in assisting prompt creation, enhancing understanding of architectural concepts, and improving prompting skills. This allowed to assess the direct impact of the interventions. Open-ended questions provided participants with the opportunity to offer qualitative feedback on the customized GPT model, including aspects they found most helpful, suggestions for improvement, challenges faced, and thoughts on the impact of AI tools in architectural education and practice.

**Section A: Self-Assessment of Prompting Skills**

*Please rate your agreement with the following statements on a scale from 1 to 5, where 1 means "Not at All" and 5 means "Extremely."*

| Question Number | Statement | Rating Scale (1-5) |
|---|---|---|
| 1 | I am confident in my ability to create effective prompts for text-to-image AI tools. | 1 - 2 - 3 - 4 - 5 |
| 2 | I understand how to structure prompts to achieve desired visual outcomes. | 1 - 2 - 3 - 4 - 5 |
| 3 | I am confident in my ability to improve AI-generated images through prompt refinement. | 1 - 2 - 3 - 4 - 5 |

**Section B: Perceived Improvement**

| Question Number | Statement | Rating Scale (1-5) |
|---|---|---|
| 4 | My prompting skills have improved as a result of this experiment. | 1 - 2 - 3 - 4 - 5 |
| 5 | My understanding of AI tools in architectural design has enhanced. | 1 - 2 - 3 - 4 - 5 |
| 6 | I am more likely to use text-to-image AI tools in my future academic or professional work. | 1 - 2 - 3 - 4 - 5 |

**Section C: Experience with AI Guides and Personas** *(For Experimental Groups Only)*

| Question Number | Statement | Rating Scale (1-5) |
|---|---|---|
| 7 | The tailored prompting guide was helpful in assisting me to create better prompts. | 1 - 2 - 3 - 4 - 5 |
| 8 | The AI personas enhanced my understanding of architectural concepts. | 1 - 2 - 3 - 4 - 5 |
| 9 | The AI personas were effective in improving my prompting skills. | 1 - 2 - 3 - 4 - 5 |

**Section D: Overall Experience**

| Question Number | Statement | Rating Scale (1-5) |
|---|---|---|
| 10 | I am satisfied with the quality of the images generated based on my prompts. | 1 - 2 - 3 - 4 - 5 |
| 11 | The immediate feedback provided on my prompts was useful. | 1 - 2 - 3 - 4 - 5 |
| 12 | The experiment enhanced my creativity in architectural design. | 1 - 2 - 3 - 4 - 5 |
| 13 | I found the experiment engaging and enjoyable. | 1 - 2 - 3 - 4 - 5 |

**Section E: Comparison of Before and After**

| Question Number | Statement | Rating Scale (1-5) |
|---|---|---|
| 14 | Compared to before the experiment, my overall experience with AI tools is now better. | 1 - 2 - 3 - 4 - 5 |
| 15 | Compared to before the experiment, my ability to create effective prompts has improved. | 1 - 2 - 3 - 4 - 5 |

**Section F: Open-Ended Questions**

| Question Number | Question | Answer Type |
|---|---|---|
| 16 | What aspects of the customized GPT (AI personas, prompting guide, feedback) did you find most helpful? | [Open Response] |
| 17 | What suggestions do you have for improving the customized GPT or the overall experience? | [Open Response] |
| 18 | Please describe any challenges you faced during the experiment. | [Open Response] |
| 19 | How do you think AI tools like this can impact architectural education and practice? | [Open Response] |
| 20 | Any additional comments or feedback: | [Open Response] |

Table 2. Post survey questions

## 4.2 Ethical Considerations

This study was reviewed and approved by the Institutional Review Board (IRB) at the Singapore University of Technology and Design (SUTD). All participants were informed of the purpose of the research and provided written informed consent prior to participation. Participation was voluntary, and individuals could withdraw at any time without consequence. Data collected from surveys and experimental tasks were anonymized and stored securely to protect participant confidentiality. The study adhered to ethical guidelines concerning human subject research, in accordance with institutional and national regulations.

# 5 Results

## 5.1 Participant demographics

A total of 36 students participated in this study, all of whom were enrolled in the Architecture and Sustainable Design (ASD) pillar at the Singapore University of Technology and Design (SUTD). The participants were distributed evenly across three experimental groups and selected using a stratified random sampling method to ensure diversity in experience, educational background, and gender representation. As shown in Figure 10, the gender distribution of participants was relatively balanced, with 56% identifying as female and 44% as male. In terms of experience in architecture or related fields, Figure 11 illustrates that 33% of participants reported having no prior professional experience, while the remaining 67% had varying degrees of experience: 19% had 1–2 years, another 19% had 2–4 years, and 28% had more than 4 years. This range ensured a meaningful comparison of outcomes across both novice and experienced participants. Regarding educational level, Figure 12 shows that participants were primarily drawn from Bachelor's (40%) and Master's (40%) programs, with PhD students comprising 20% of the sample. This composition allowed the study to examine how AI-assisted prompting strategies performed across different academic maturity levels, from early-stage learners to advanced researchers.

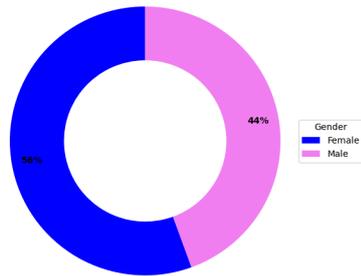

Figure 10. Gender distribution of participants.

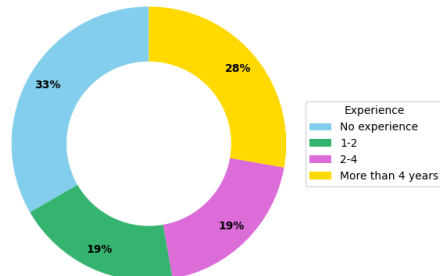

Figure 11. Participants' prior experience.

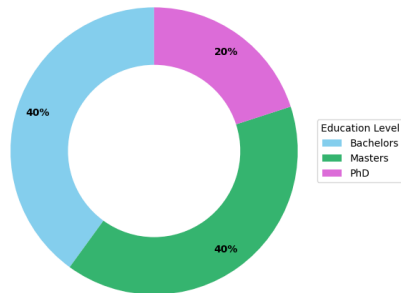

Figure 12. Educational background of participants.

## 5.2 Pre-Experiment Survey Results

Prior to beginning the prompting tasks, all 36 participants completed a pre-experiment survey designed to evaluate their familiarity with AI tools and their confidence in using text-to-image platforms relevant to architectural design. The survey results, visualized in Figure 13, provide insight into the participants' baseline competencies and attitudes toward AI in design workflows. Overall, the distribution of responses suggests a moderately experienced cohort, with most answers concentrated around the mid-range values of 3 and 4 across the nine Likert-scale questions.

In the first three questions, which addressed general experience with AI tools and confidence in adapting to new ones, the majority of participants reported moderate familiarity. Responses in Group 1, the control group, leaned more conservatively toward level 3, whereas Groups 2 and 3—those who would later interact with prompting guides and AI personas—showed a slightly higher tendency toward level 4. This pattern suggests that participants in the experimental groups may have entered the study with slightly greater self-assessed confidence or enthusiasm toward engaging with AI-assisted design tasks. When asked specifically about their use of text-to-image AI tools (Question 5 and Question 6), most participants across all groups indicated a lack of extensive prior use. Group 1, in particular, showed a larger proportion of responses clustered at level 2 and 3, highlighting their limited exposure to these tools. However, Group 3 displayed a more even spread of responses, including higher ratings at level 4 and 5, indicating some familiarity with such platforms prior to the experiment. This suggests that although the sample overall was relatively new to prompt-based image generation, certain participants—especially those in the guided groups—may have had marginally more experience or willingness to engage with this technology. Confidence in generating and refining effective prompts (Questions 7 through 9) was also moderate. Group 1 exhibited greater uncertainty, with lower frequencies of level 4 and 5 responses, suggesting limited self-assurance in their ability to craft and improve AI-generated images. In contrast, participants in Group 2 and Group 3 demonstrated a slightly more optimistic outlook, with a higher concentration of responses toward the upper end of the scale. This trend could indicate a pre-existing advantage or greater adaptability in those groups, possibly influencing their subsequent performance during the experiment.

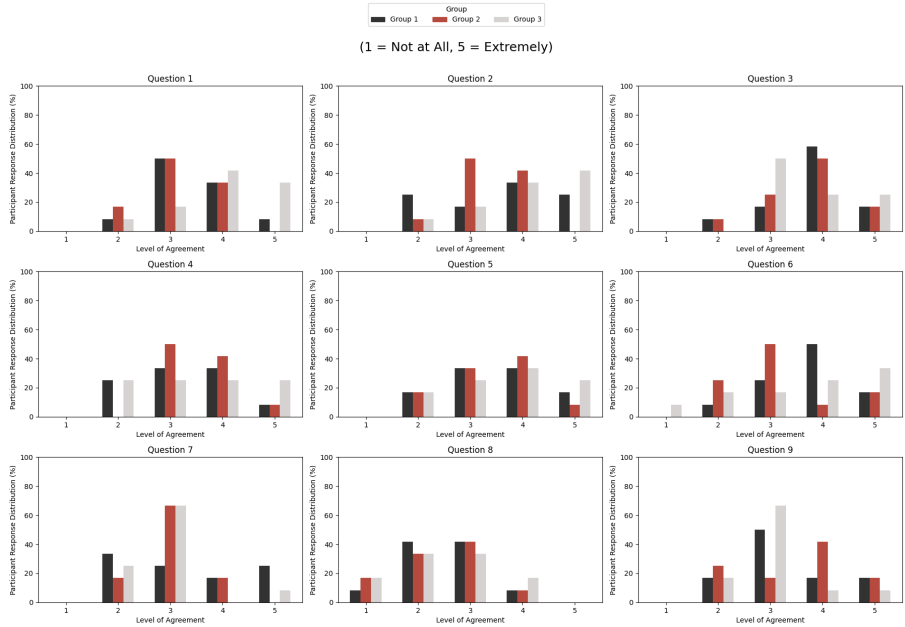

Figure 13. Pre survey results.

## 5.3 Experimental tasks and prompt performance analysis
### 5.3.1 Word Count

Figure 14 reveals a clear gradient across the three groups. Group 1, which had access to both the prompting guide and AI personas, consistently produced the longest prompts, with word counts increasing from 27 to 54 across tasks. Group 2, assisted by the prompting guide only, followed closely, reaching a maximum of 43 words in the final task. In contrast, Group 3, the control group with no AI assistance, consistently wrote the shortest prompts, starting with just 15 words and peaking at 41. This pattern strongly suggests that both the guide and AI personas encouraged participants to elaborate more fully, using richer and more varied architectural vocabulary. The AI support tools likely helped participants better understand how to structure detailed prompts, and gave them confidence to describe images more thoroughly.

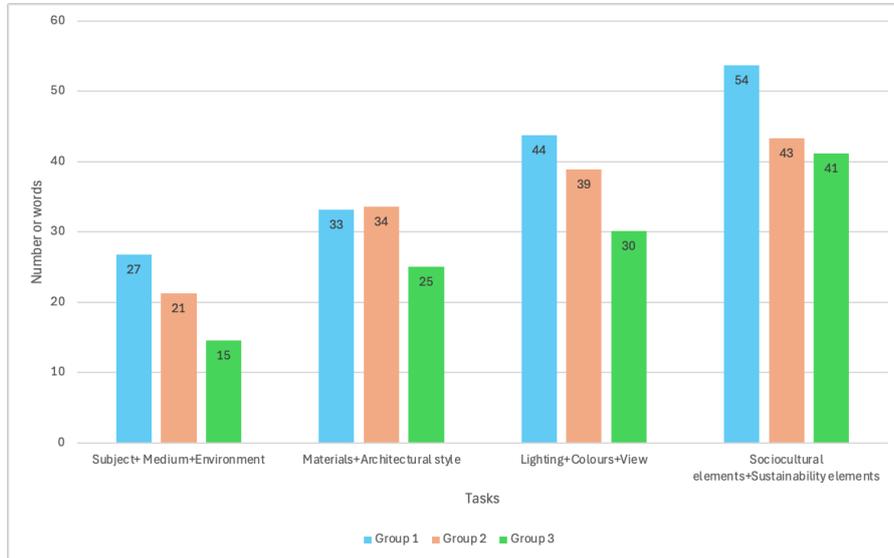

Figure 14. Average word count in prompts per task across experimental groups

A one-way ANOVA (see Table 3) revealed a statistically significant difference in word count between the groups, $F(2, 141) = 4.45$, $p = .013$. Post hoc analysis using Tukey's HSD (see Table 4) indicated that the difference between Group 1 and Group 3 was statistically significant, with Group 1 demonstrating a notably higher mean word count. The difference between Group 2 and Group 3 was present but less pronounced, and no significant difference was found between Group 1 and Group 2. These results suggest that AI personas and structured prompting guides are effective in encouraging participants to write more elaborate and detailed prompts. The increased word count in the experimental groups may reflect improved prompting literacy and confidence in architectural vocabulary, while the brevity in the control group points to either limited prompting skills or lower engagement with the descriptive requirements of the tasks. The use of AI-supported scaffolding thus appears to enhance both prompting depth and linguistic expansion in architectural design contexts.

| Source | Sum of Squares | df | Mean Square | F-value | p-value |
|---|---|---|---|---|---|
| Group | 2087.18 | 2 | 1043.59 | 4.45 | 0.013 |
| Residual | 33085.98 | 141 | 234.65 | | |

Table 3. One-way ANOVA results for word count across experimental groups.

| group1 | group2 | meandiff | p-adj | lower | upper | reject |
|---|---|---|---|---|---|---|
| Group1 | Group2 | -5.0833 | 0.2382 | -12.49 | 2.3233 | FALSE |
| Group1 | Group3 | -9.3125 | 0.0095 | -16.7191 | -1.9059 | TRUE |
| Group2 | Group3 | -4.2292 | 0.3687 | -11.6358 | 3.1775 | FALSE |

Table 4. Tukey HSD post hoc comparison of word count between experimental groups.

### 5.3.2 Time Spent

Time spent on prompt creation shows a parallel trend, see Figure 15. Group 1 again led with the longest average time per task, culminating at 5.1 minutes for the final and most complex task. Group 2 maintained a slightly shorter but still engaged average, while Group 3 completed tasks much faster, with a maximum of 4.0 minutes. While speed might appear beneficial, in this context it likely reflects a lack of depth or uncertainty. The longer time spent by Group 1 indicates that participants were more engaged in thoughtful prompt construction, likely drawing on the support of AI personas to experiment with vocabulary and structure. Group 2's results suggest that even the guide alone encouraged more deliberate engagement compared to the unguided group 3.

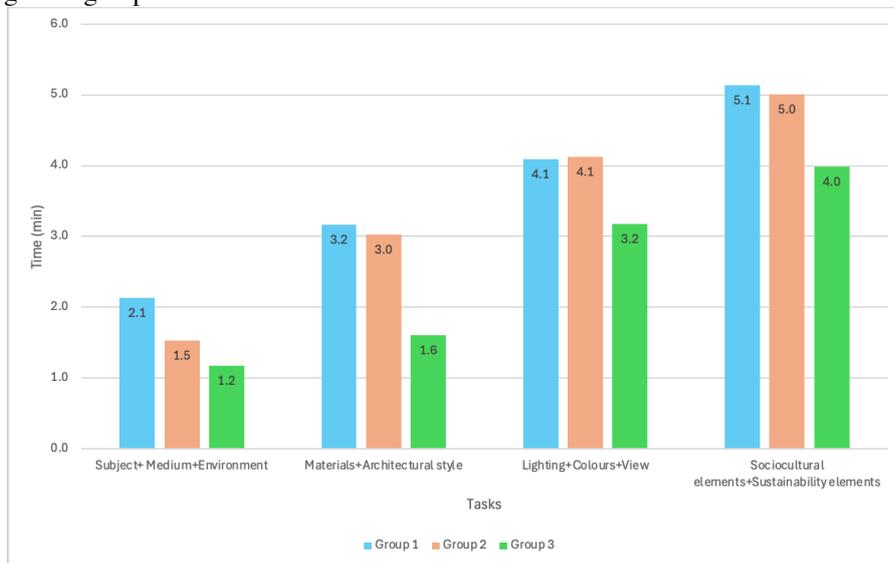

Figure 15. Average time spent in prompts per task across experimental groups.

To assess how the availability of AI support influenced participants' engagement and effort, an analysis was conducted on the time spent writing prompts across all tasks and groups. While Group 1 participants spent the most time per task on average, Group 2 demonstrated moderate time investment, and Group 3 completed their prompts the fastest. A one-way ANOVA (see Table 5) revealed that these differences in time were not statistically significant, $F(2, 141) = 2.65$, $p = .074$. Although the result did not reach the conventional threshold of significance ($p < .05$), the proximity of the p-value suggests a potential trend toward increased time investment when participants are provided with AI support tools. A follow-up Tukey HSD post hoc test (see Table 6) further confirmed that none of the pairwise differences between groups were statistically significant. However, the largest numerical difference was observed between Group 1 and Group 3, aligning with the descriptive findings. Participants in Group 1 appeared to take more time to reflect and construct richer prompts, likely influenced by the additional feedback and vocabulary suggestions provided by the AI personas. In contrast, the shorter times recorded for Group 3 may reflect more intuitive or minimal prompting behavior, possibly due to limited prompting literacy or confidence. While not statistically conclusive, these patterns suggest that AI-guided prompting tools may encourage deeper cognitive engagement, leading participants to spend more time refining their inputs, a finding consistent with trends observed in the word count and similarity analyses.

| Source | Sum of Squares | df | Mean Square | F-value | p-value |
| --- | --- | --- | --- | --- | --- |
| Group | 23.94 | 2 | 11.97 | 2.65 | 0.074 |
| Residual | 636.02 | 141 | 4.51 | | |

Table 5. One-way ANOVA results for time spent across experimental groups.

| group1 | group2 | meandiff | p-adj | lower | upper | reject |
| --- | --- | --- | --- | --- | --- | --- |
| Group1 | Group2 | -0.2146 | 0.8738 | -1.2415 | 0.8123 | FALSE |
| Group1 | Group3 | -0.9521 | 0.0753 | -1.979 | 0.0748 | FALSE |
| Group2 | Group3 | -0.7375 | 0.2083 | -1.7644 | 0.2894 | FALSE |

Table 6. Tukey HSD post hoc comparison of time spent between experimental groups.

### 5.3.3 Similarity

The similarity scores reflect how closely participant-generated prompts matched the original prompts generated by the customized GPT, see Figure 16. Once again, Group 1 performed best, with similarity scores ranging from 72% to 82%, indicating that the AI guidance helped participants align more closely with optimal prompting language. Group 2 followed with moderately high scores, while Group 3 consistently showed the lowest similarity, particularly in tasks involving more complex parameters (e.g., "Materials and Architectural Style" and "Sociocultural and Sustainability elements"), with scores as low as 58%. These findings imply that without structured support, participants in Group 3 struggled to identify and replicate the nuances expected in architectural prompt construction.

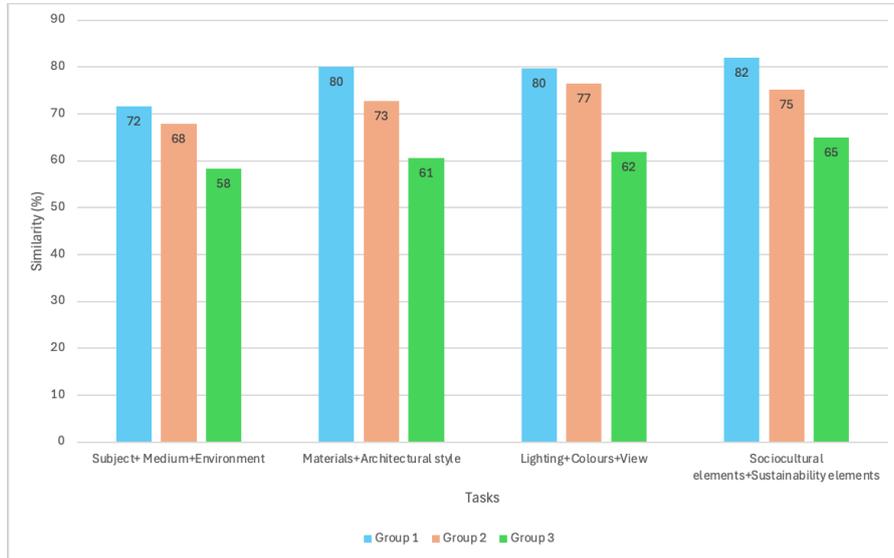

Figure 16. Average similarity percentage in prompts per task across experimental groups.

A one-way ANOVA (see Table 7) revealed a statistically significant difference in similarity scores between the groups, $F(2, 141) = 11.63$, $p < .001$. This suggests that the type of support provided, whether through AI personas, a structured prompting guide, or no assistance had a measurable impact on participants' ability to match the reference prompt structure. Tukey HSD post hoc (see Table 8) comparisons confirmed that Group 1 (provided with both prompting guide and AI personas) and Group 2 (prompting guide only) achieved significantly higher similarity scores than Group 3, the control group that received no support. The difference between Group 1 and Group 2, however, was not statistically significant. These findings indicate that access to a prompting guide regardless of whether AI personas were also used, substantially improved participants' ability to generate structurally aligned and semantically relevant prompts. In contrast, participants without support (Group 3) demonstrated greater divergence from the target format, likely due to a lack of vocabulary scaffolding or prompt organization strategies. Overall, the results reinforce the effectiveness of guided prompting tools in improving accuracy and alignment in AI-supported architectural design tasks, and suggest that even lightweight instructional scaffolding can significantly enhance user performance.

| Source | Sum of Squares | df | Mean Square | F-value | p-value |
|---|---|---|---|---|---|
| Group | 3328.63 | 2 | 1664.31 | 11.63 | < .001 |
| Residual | 20181.38 | 141 | 143.13 | | |

Table 7. One-way ANOVA results for similarity percentage across experimental groups.

| group1 | group2 | meandiff | p-adj | lower | upper | reject |
|---|---|---|---|---|---|---|
| Group1 | Group2 | -5.1875 | 0.0886 | -10.9721 | 0.5971 | FALSE |
| Group1 | Group3 | -11.75 | 0.0 | -17.5346 | -5.9654 | TRUE |
| Group2 | Group3 | -6.5625 | 0.0219 | -12.3471 | -0.7779 | TRUE |

Table 8. Tukey HSD post hoc comparison of similarity percentage between experimental groups.

### 5.3.4 Concreteness

Concreteness scores reveal how literal or abstract participants' prompts were, see Figure 17. While all groups maintained reasonably high scores, Group 1 and Group 2 slightly outperformed Group 3, particularly in the more complex tasks. Group 1 peaked at 3.9 to 4.0 in tasks involving lighting, sociocultural elements, and sustainability, likely due to the vocabulary suggestions and thematic depth provided by AI personas. Group 3, on the other hand, remained more general and less descriptive, averaging 3.4 to 3.7. This suggests that participants without support tended to use more generic or less specific language, possibly due to uncertainty or limited prompting literacy.

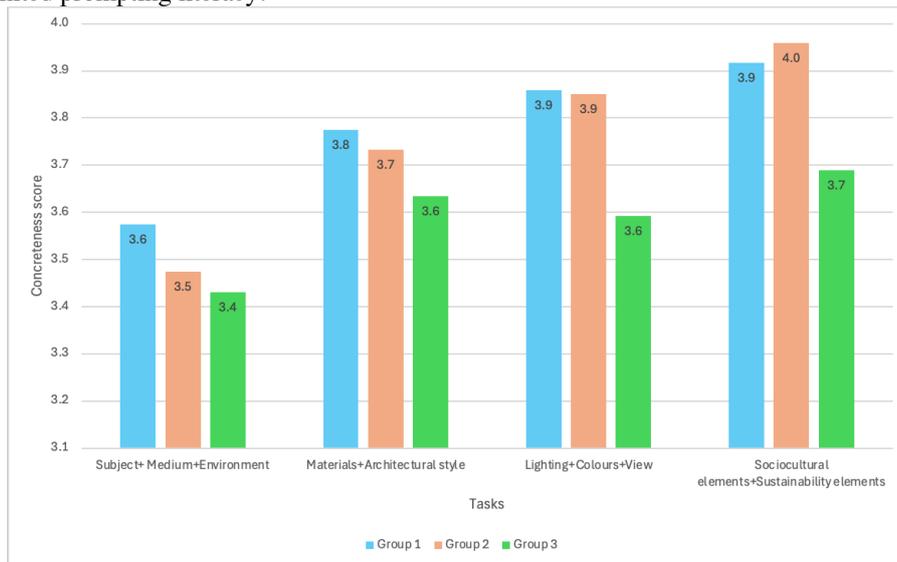

Figure 17. Average concreteness score in prompts per task across experimental groups.

A one-way ANOVA (see Table 9) was conducted to determine whether there were significant differences in concreteness scores among the three experimental groups. The analysis revealed that the differences were not statistically significant, $F(2, 141) = 1.85$, $p = .161$. This indicates that the presence or absence of prompting guidance and AI personas did not produce measurable differences in how concrete participants' prompts were. Post hoc comparisons using Tukey's HSD test (see Table 10) confirmed this result, as no statistically significant pairwise differences were found between any of the groups. Although descriptive data showed slightly higher concreteness scores in Group 1 and Group 2 compared to Group 3, these differences were not large enough to reach significance. These findings suggest that while AI-based tools were effective in improving other aspects of prompt quality, such as word count and similarity, they did not significantly alter the literalness or specificity of language used. This may be due to the fact that participants across all groups were architecture students, who already possessed a baseline ability to describe physical and material characteristics with relative clarity, regardless of intervention.

| Source | Sum of Squares | df | Mean Square | F-value | p-value |
|---|---|---|---|---|---|
| Group | 0.461 | 2 | 0.230 | 1.85 | 0.161 |
| Residual | 17.547 | 141 | 0.124 | | |

Table 9. One-way ANOVA results for concreteness score across experimental groups.

| group1 | group2 | meandiff | p-adj | lower | upper | reject |
|---|---|---|---|---|---|---|
| Group1 | Group2 | -0.0271 | 0.925 | -0.1977 | 0.1435 | FALSE |
| Group1 | Group3 | 0.1042 | 0.32 | -0.0664 | 0.2747 | FALSE |
| Group2 | Group3 | 0.1313 | 0.1659 | -0.0393 | 0.3018 | FALSE |

Table 10. Tukey HSD post hoc comparison of concreteness score between experimental groups.

**5.4 Correlation analysis**

To explore the relationships between the main prompt performance metrics, a correlation analysis was conducted between time spent, word count, similarity, and concreteness scores across all participants.

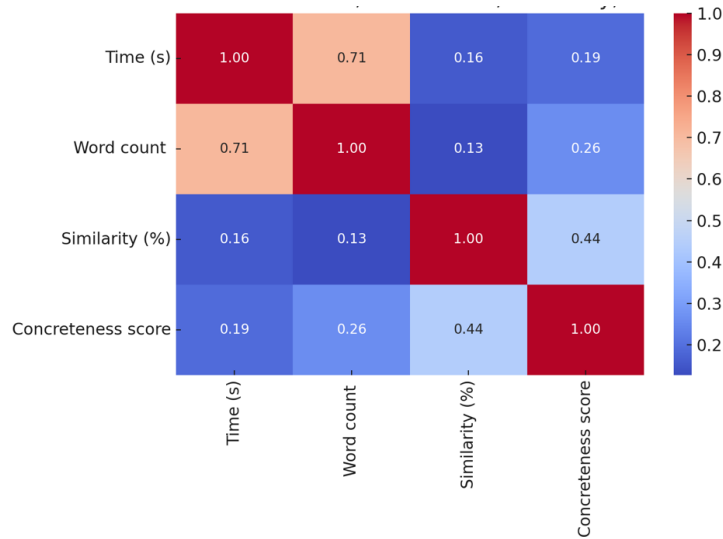

Figure 18. Correlation matrix between time, word count, similarity, and concreteness across all groups.

The overall correlation matrix (see Figure 18) revealed several key trends. A strong positive correlation was observed between time and word count ($r = 0.71$), indicating that participants who spent more time on a task tended to produce longer prompts. A moderate positive correlation was also found between similarity and concreteness ($r = 0.44$), suggesting that prompts written with more concrete and literal language were more likely to align with the GPT-generated reference prompts. Additionally, word count and concreteness showed a mild positive correlation ($r = 0.26$), while other relationships (e.g., time and similarity) were weaker and less consistent. To better understand the effect of AI support, correlation matrices were also analyzed by group (see Figure 19):

- Group 1 exhibited a very strong correlation between time and word count ($r = 0.86$), but negative correlations between time/word count and similarity ($r = –0.29$ and $–0.41$, respectively). This suggests that, despite spending more time and writing more words, participants may have deviated from the reference structure—possibly reflecting more exploratory or creative prompting. However, similarity and concreteness had a very strong positive correlation ($r = 0.75$), showing that accurate prompts were also the most literal and specific.
- Group 2 showed moderate-to-strong correlations among word count, similarity, and concreteness ($r = 0.38$–$0.60$). Notably, word count and concreteness were more tightly linked ($r = 0.59$), suggesting that the structured guide helped participants compose longer prompts using more specific vocabulary. Time, however, was only weakly correlated with these metrics.
- Group 3 also showed a moderate relationship between word count and similarity ($r = 0.60$) and between similarity and concreteness ($r = 0.58$). These values indicate that, even without external scaffolding, participants who wrote longer prompts tended to produce more accurate and concrete ones. However, all correlations in this group were lower compared to Group 1 and Group 2, highlighting the role of AI support in reinforcing these connections.

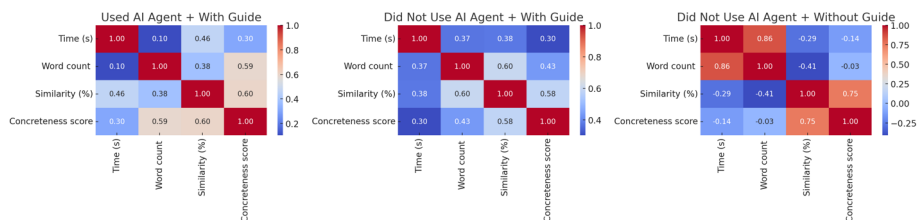

Figure 19. Correlation matrices between performance variables by experimental group.

Overall, the analysis suggests that AI-driven tools like prompting guides and personas affect not only the individual performance metrics but also the way these metrics interrelate. Specifically, while unguided participants showed natural correlations between clarity and accuracy, those using AI tools demonstrated more complex patterns, sometimes trading off direct alignment with the GPT prompt in favor of deeper or more interpretive prompting. These findings underscore the need to assess not just output quality but the cognitive processes and prompt strategies shaped by AI assistance.

**5.5 Post-Experiment Survey Results**

Following the completion of the experimental tasks, participants completed a post-survey that assessed their confidence, perceived improvement, and overall experience using AI tools for architectural prompting. As illustrated in Figure 20, the responses across all groups skewed strongly toward the upper end of the Likert scale, indicating high levels of satisfaction and self-reported skill development. For Questions 1 to 3, which evaluated participants' confidence in crafting and refining prompts, the majority of responses were concentrated at level 4 and 5 across all groups, with a notable increase in Group 3, the cohort that worked with both the prompting guide and AI personas. In these questions, Group 3 consistently showed a higher distribution of "extremely confident" responses compared to Group 1, suggesting that the presence of AI personas provided valuable support.

Questions 4 to 6 focused on perceived improvements in prompting skills and the intention to use text-to-image tools in the future. All three groups exhibited an upward shift in agreement levels compared to their pre-survey responses. Group 2 and Group 3 showed particularly high agreement in Question 4 (prompting skill improvement), with over 70% of participants in Group 3 selecting level 5, suggesting a strong impact of the AI personas and structured guidance. Similarly, Questions 7 to 9, which were only answered by participants in the experimental groups, revealed that both the prompting guide and AI personas were widely regarded as helpful. Most responses landed at level 5, reinforcing the idea that these tools contributed to both a deeper understanding of architectural concepts and an improved ability to generate prompts. The remaining questions (10 to 15) evaluated participants' overall satisfaction with the experiment, the feedback system, and their comparative experience before and after participation. The results here further confirm the intervention's effectiveness. Nearly all participants rated the feedback mechanism and their engagement with the experiment at level 4 or 5. Of particular interest is Question 14, where participants compared their AI experience before and after the experiment. A clear shift toward higher confidence was seen across all three groups, with Group 3 showing the most dramatic improvement—a trend that supports the claim that layered support systems (both guides and AI personas) enhance learning outcomes in complex design tasks.

When comparing these post-survey results with the earlier pre-survey data, the contrast is striking. In the pre-survey, most participants hovered around levels 3 and 4 when asked about their ability to craft or refine prompts (e.g., Questions 1, 6, and 9 from the pre-survey). In the post-survey, those same questions received responses mostly at levels 4 and 5, especially from participants in the experimental groups. The largest gain is evident in Group 1, where the combination of structured tips and expert-like AI personas led to more pronounced growth in confidence, understanding, and engagement. This outcome suggests that the integration of guided support mechanisms significantly enhances not only participants' technical prompting skills but also their motivation and readiness to incorporate AI tools into their architectural practice.

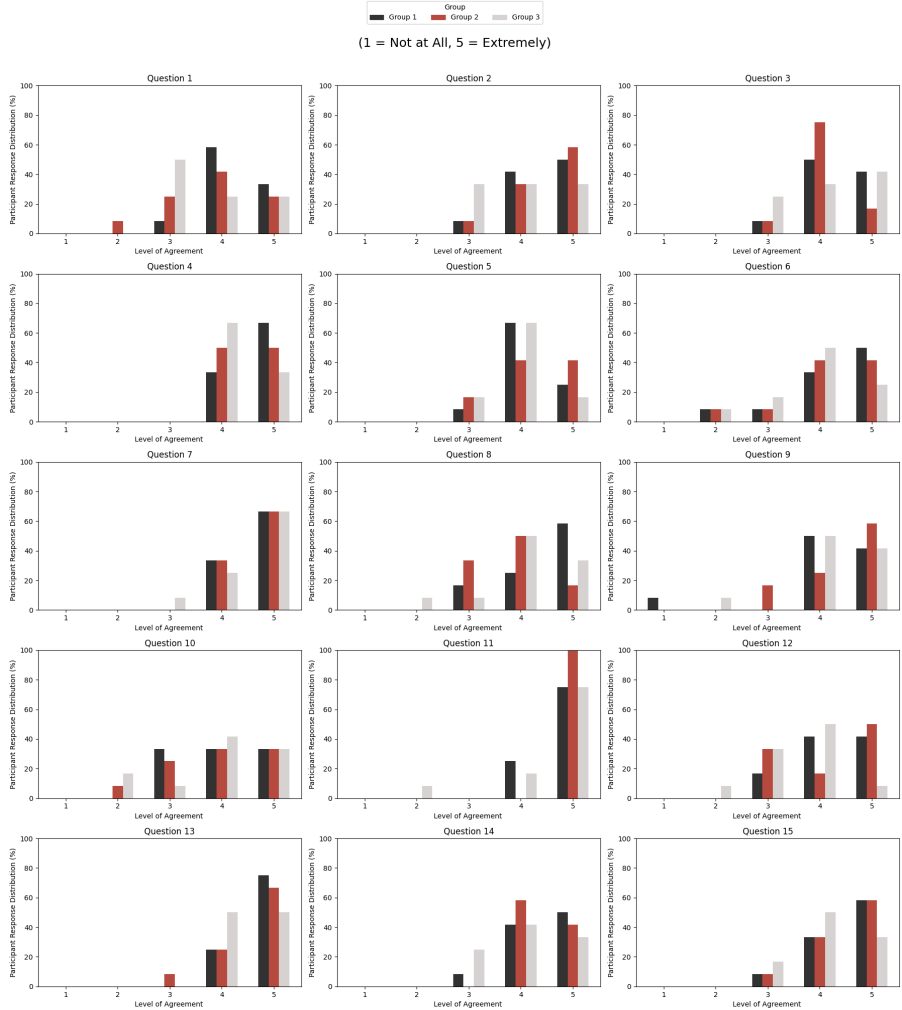

Figure 20. Post survey results.

Open-ended responses from participants offer valuable insights into their experience using the customized GPT system during the experiment. Overall, the most frequently praised features included the prompting guide, step-by-step feedback, and the scoring system that helped participants refine their architectural vocabulary and understand how their prompts aligned with intended design outputs. Many participants described the process as a "great learning experience," citing that the guidance on lexical choices and prompt structure helped improve both their confidence and precision. Several responses emphasized that the comparison between their own prompt and a model prompt made the learning more tangible and effective, particularly for those unfamiliar with AI's interpretive quirks. Participants also appreciated the AI personas, especially the architect and sustainability consultant personas, which were viewed as helpful in expanding vocabulary and encouraging interdisciplinary thinking. Some noted that the feedback loop helped them recognize that prompts which seemed clear from a human perspective were often too vague for AI to process meaningfully. This led to a better understanding of how to "speak to AI," a concept many found eye-opening and directly transferable to their future work.

Despite these strengths, participants also offered constructive suggestions. A common request was for a clearer explanation of prompting parameters, such as what constitutes a "medium" or "lighting style," especially for non-native English speakers or those unfamiliar with architectural terminology. Others recommended adding databases of sample keywords, real-time word count indicators, and more visual previews for prompt components—such as styles or materials—to improve usability. Several users also expressed a desire for more culturally diverse image outputs, mentioning that the model leaned heavily toward Eurocentric or East Asian architectural aesthetics. A few responses highlighted technical limitations such as low image resolution or the difficulty of inferring conceptual design elements from images that emphasized only physical features.

Challenges faced by participants ranged from vocabulary limitations and word count constraints to confusion with AI feedback, especially when suggestions were perceived as inaccurate or misaligned with the original prompt intention. For non-native speakers, language clarity and translation inconsistencies were also recurring issues, as were cases where participants felt their background knowledge was insufficient to articulate complex architectural concepts.

## 6. Discussion

The results reveal that participants in Group 1, who had access to both support tools, consistently produced longer, more accurate, and more specific prompts than their peers. These outcomes align with research showing that layered scaffolding—combining static instruction with dynamic feedback—enhances learning performance, particularly in complex cognitive tasks (Collins et al., 1989; Roll et al., 2014). The significant gains in word count and similarity scores for Group 1 suggest that the dual-support strategy enabled deeper elaboration and clearer articulation of architectural intent. The structured guide served as a cognitive framework, while AI personas functioned as interactive tutors offering domain-specific vocabulary, effectively simulating interdisciplinary collaboration (Han et al., 2024). These findings reflect the value of "cognitive apprenticeship," where learners benefit from expert modeling, feedback, and progressive autonomy (Collins et al., 1989; Holmes et al., 2021).

Interestingly, concreteness scores did not significantly differ among groups, indicating that students already possessed foundational skills in describing architectural elements. This supports prior work noting that architectural education inherently develops the ability to communicate physical and spatial characteristics (Oxman, 2008). The minimal change in concreteness may reflect a ceiling effect (Salkind, 2010), where instructional interventions have limited impact on skills that are already well-developed. A notable pattern emerged in the correlation analyses: in Group 1, longer time spent and higher word counts were negatively correlated with similarity scores. This suggests that while support tools fostered more detailed and creative prompting, they may also have encouraged divergence from model answers. This trade-off between creative exploration and model adherence mirrors the "exploration-exploitation" dilemma in design thinking (Gero & Maher, 2013), and highlights the importance of balancing freedom and structure in AI-assisted learning environments (Jansson & Smith, 1991).

The structured guide alone (Group 2) also yielded measurable improvements, particularly in word count and vocabulary usage, though not to the same extent as the combined support in Group 1. This confirms that even light scaffolding can significantly support novice learners (Vygotsky, 1978), but richer, interactive guidance provides additional pedagogical value. These findings advocate for multi-tiered instructional designs in architecture education, where static prompts are reinforced with dynamic, responsive tools (Zhang et al., 2023). The feedback system played a crucial role in facilitating self-regulated learning. By visualizing performance metrics such as similarity and concreteness, students gained insights into the clarity and quality of their prompts. Prior work has shown that timely, targeted feedback enhances metacognitive awareness and fosters a reflective learning cycle (Shute, 2008; Nicol & Macfarlane-Dick, 2006). Open-ended survey responses further confirmed that students valued the ability to compare their prompts to model outputs and iteratively refine their work.

Prior AI experience played a minor role in determining performance. While students with more experience began with slightly higher confidence levels, both experimental groups exhibited significant improvement, demonstrating the accessibility of the tools for novices. This reflects findings by Long and Magerko (2020), who noted that well-designed scaffolds can rapidly build AI literacy regardless of initial skill levels. From a design perspective, the study underscores key features for future AI-enhanced learning environments: (1) modular scaffolding that transitions from structured guides to open-ended interaction; (2) clear, explainable feedback mechanisms; and (3) persona-driven interdisciplinary support. Tools should also account for inclusivity—students suggested more culturally diverse visual outputs and clearer definitions of prompt parameters—echoing broader calls for responsible AI design (Holmes et al., 2021).

Nonetheless, challenges were identified. Some students found AI feedback ambiguous or misaligned with their design intentions. Similar issues have been observed in AI-supported writing, where unclear suggestions often reduce user trust (Lee & Dey, 2023). Future systems should improve explainability by attaching rationale to AI suggestions and allowing two-way interaction where students can query the feedback (Weller et al., 2020).Educators must be cautious not to let AI tools over-direct students. While support structures are essential for novices, excessive reliance can inhibit creativity and autonomy (Lee et al., 2022). Fading scaffolds and reflective tasks—such as attributing decisions to either the AI or the student—can help preserve ownership and foster creative confidence (Collins et al., 1989).

This study is limited by its relatively small sample size and single-institution context, which may affect generalizability. All participants were architecture students already trained in design description, possibly contributing to the high baseline concreteness. Future research should explore how these tools perform with less experienced users, such as first-year students or learners from other design disciplines. Further studies could investigate long-term impacts of AI-supported prompting on design outcomes, professional communication, and studio critique performance. Exploring adaptive AI personas that respond to student progress over time, or integrating these tools into collaborative studio workflows, could yield richer insights. Additionally, incorporating culturally inclusive training data and multilingual support would broaden accessibility and relevance across global contexts.

## 7 Conclusions

This study evaluated the integration of customized GPT models and structured prompting guides into architectural education, specifically aiming to enhance students' prompting proficiency when interacting with AI-generated image platforms. The results confirmed the effectiveness of incorporating structured prompting guides and AI personas into architectural curricula.

Quantitative analyses demonstrated statistically significant improvements in key performance metrics, particularly in word count and prompt similarity. Participants in the experimental groups, especially those supported by AI personas and structured guides generated significantly longer, more detailed prompts compared to the control group. The similarity scores indicated these participants aligned their prompts more closely with optimal architectural language and format. These outcomes strongly suggest that tailored GPT interactions enhance students' prompting literacy, enabling clearer and more effective communication of architectural concepts.

The correlation analysis provided further insights, showing that participants with AI-driven support spent more time on tasks and generated longer prompts. Interestingly, while greater time and word count correlated negatively with prompt similarity in the AI-supported group indicating a potential trade-off between creative exploration and strict adherence to reference prompts, these participants also showed the highest scores in concreteness and specificity. Thus, AI-driven tools not only encouraged more detailed descriptions but also fostered deeper cognitive engagement and exploration of architectural vocabulary.

Qualitative feedback from participants reinforced these findings, highlighting increased confidence and prompting skills due to the interactive and iterative feedback mechanisms provided by AI personas and structured guides. Participants valued the immediate feedback and tailored vocabulary suggestions, emphasizing the practical utility of these tools in educational settings. However, they also identified areas for improvement, including clearer definitions of prompting parameters and increased cultural diversity in generated outputs.

Despite these promising results, several limitations were noted. For instance, although AI support improved the quantity and quality of prompts, the negative correlation between time and similarity suggests a potential need for balancing creativity with structured guidance. Additionally, the concreteness scores did not significantly differ across groups, indicating that all participants, regardless of intervention, possessed a foundational skill set for describing architectural elements clearly. Future research could address these limitations by testing methods to optimize the balance between creativity and precision, exploring more targeted interventions to enhance concreteness, and expanding the diversity of participants. Further studies could also investigate long-term impacts of AI-supported prompting on architectural education outcomes and professional practice. This research demonstrates the substantial pedagogical benefits of integrating tailored AI-driven tools into architectural curricula, providing a robust basis for continued exploration and refinement.

**Disclosure of interests.** The authors have no competing interests to declare that are relevant to the content of this article.

**Credit author statement.** Juan David Salazar Rodriguez: Conceptualization, Methodology, Investigation, Validation, Data curation, Writing – Original Draft, Visualization. **Sam Conrad Joyce:** Conceptualization, Supervision, Writing – Review & Editing. **Julfendi Julfendi:** Software, Validation, Writing – Review & Editing.